\documentclass[11pt,twoside]{article}
\usepackage{newpasp}

\def\edcomment#1{\iffalse\marginpar{\raggedright\sl#1\/}\else\relax\fi}
\reversemarginpar

\begin{document}
\title{Acceleration and Dissipation in Relativistic Winds}
 \author{Jonathan Arons}
\affil{University of California, Berkeley, Department of Astronomy,
601 Campbell Hall, Berkeley, CA 94720-3411, USA}

\begin{abstract}
I argue that ideal MHD relativistic winds are always limited 
in practice to asymptotic 4-velocity
$\gamma_\infty \approx \sigma_0^{1/3}$ and asymptotic magnetization 
$\sigma \sim \sigma_0^{2/3} \gg 1$, where $\sigma_0$ is the wind magnetization
with respect to the rest energy density, evaluated at the light cylinder of
the rotating, magnetized compact object that drives the flow.  This suggests that
the observed low value of he asymptotic $\sigma$ in the equatorial sectors of
the winds driving Pulsar Wind Nebulae and the associated high values of the asymptotic
4-velocity are a consequence of magentic dissipation in the wind zone.
\end{abstract}

\section{Introduction}

Pulsar Wind Nebulae (PWNe) provide our nearest at hand examples of 
relativistic outflows from compact objects. They also form incontrovertible 
examples of electromagnetically driven flows. Thus their physics is of interest,
both in its own right and as purveyors of insight into more remote systems known
to have relativistic outflow dynamics, such as AGN and GRB jets, and suspected 
of being electromagnetically driven.

\section{Observations of PWNe Relevant to Relativistic Wind Physics}

Observations and models of  young PWNe allow
us to diagnose the properties of the relativistic winds
conveying the underlying compact objects' rotational energy to their nebulae. 
1D flow models applied to the Crab Nebula 20 years ago by Kennel and Coroniti 
(1984a,b) led to the conclusion that the {\em equatorial} relativistic wind from 
the Crab pulsar just upstream of the wind's termination shock has some 
surprising properties: 
a terminal Lorentz factor $\gamma_1 \approx 10^6$ and a terminal magnetization
$\sigma_1 = B_1^2/4\pi \gamma_1 \rho_1 c^2  \approx 3 \times 10^{-3} \ll 1$. 
Such models also lead to a measurement
of the $e^\pm$ injection rate ($\dot{N}_\pm \sim 10^{38.5}$ s$^{-1}$ for the Crab)
which is in reasonable accord with previous inferences from applications of 
elementary synchrotron theory to nebular X-ray emission.  Similar inferences 
have been drawn for other
PWNe (Gaensler et al. 2002; but see Reynolds 2003). Modeling (Spitkovsky
\& Arons 2003) of the time
dependent structure observed in the shock termination layer (Hester et al. 2002)
comes up with very similar values for $\gamma_1$, $\sigma_1$ and $\dot{N}_\pm $.
Such models make the additional suggestion that the equatorial wind has a high energy
ion component with energy per particle $E_i \approx q_i \Phi/m_{eff} c^2$, where
$\Phi$ is the rotational voltage of the pulsar. The inferred ion injection
rate $\dot{N}_i$ is about equal to
the electric current per unit charge expected from electrodynamic theories
of magnetized compact object spindown (the Goldreich-Julian current).  
Here $m_{eff} = m_i + 2\kappa_\pm m_\pm$ is the rest mass per particle of the
plasma in the relativistic wind, with $\kappa_\pm $ the number of pairs
per Goldreich-Julian current carrier expelled from the underlying star. 
Under
pulsar conditions, $m_i = m_p$ or $m_{He}$ and $m_{eff} = (5-10)m_p $. Such models
fail to account for the higher average particle flux required to explain the
Crab Nebula's radio emission; see Arons (1998) and Gallant et al. (2002)
for discussions of this issue.

\section{Wind Acceleration Theory}

The high wind
4-velocity $\gamma_1$ and the low magnetization $\sigma_1$ generate problems
for the basic physics of relativistic MHD wind flow; the $\dot{N}_\pm$ problem
may point to troubles with the theory of the plasma source. The existing
theory of MHD wind flow (Michel 1969, Goldreich and Julian 1970, Beskin et al 1998),
applied to rotators with split monopole poloidal magnetic field geometry, suggest
that the wind 4-velocity at ``infinity'' is $\gamma_\infty \approx \sigma_0^{1/3}$,
where $\sigma_0 \approx B_L^2 /4\pi \rho_L c^2 \approx e\Phi/2m_{eff} c^2$. 
Contopoulos et al. (1999) show that at least for the aligned rotator, the asymptotic
wind structure ($ r > $ a few $R_L = c/\Omega$) is reasonably well approximated 
by outflow in (split) monopole magnetic geometry, which is almost radial, and 
Bogovalov's (1999) generalization of split monopole outflow to oblique rotator 
geometry suggests that almost radial outflow has general applicability.
Ideal MHD wind theory applied to more general asymptotic flow structures
also suggests almost radial flow outside
the light cylinder (Begelman and Li 1994, Beskin et al 1998),
in which case $\sigma_{wind} = \sigma_0 /\gamma_{wind} \gg 1$ so long as
$\gamma_{wind} \ll \sigma_0 $. The traditional MHD solutions in monopole geometry
with $\gamma_{wind}(r) < \sigma_0^{1/3}$ thus imply 
$\sigma_{wind}(r) > \sigma_0^{2/3}$.
Everything
that has appeared on pulsar magnetosphere theory suggests $\sigma_0 \gg 1$ - for
the Crab, pair creation models (either polar cap or outer gap) suggest 
$\sigma_0 \sim 10^4$. Thus standard theory says the wind should be ``high sigma'',
all the way to the termination shock, in flagrant contradiction of the inferred
magnetization.

The origin of this theoretical prediction is interesting.   Magnetic pressure, in the form of a wound up
magnetic ``spring'', provides the driving force underlying outflows driven by
large scale electromagentic fields.  The magnetic spring can be an 
effective accelerator only so long as the magnetic field remains dynamically 
coupled to the underlying rotator - by Newton's third law, the magnetic field
has to push in against a heavy weight (the compact object), in order to exert
an outward push on the wind's plasma. Such coupling requires the field at large
radii to be able to communicate with the field's roots in the star, which means
that acceleration persists only so long as the plasma velocity is less than 
the magnetosonic speed.

Non-relativistically, such good contact translates into the
requirement $\rho v^2 < B^2 /4\pi$, where $\rho$ is the rest mass density,
or $v < v_{ms} = B/\sqrt{4\pi \rho}$ = nonrelativistic magnetosonic speed.  
Naively, one might expect that the relativistic version of this constraint
would be $\rho \gamma_{wind} c^2 < B^2/4\pi $, or 
$v_{ms} \rightarrow B/\sqrt{4\pi \gamma_{wind} \rho}$ in the relativistic case. 
With $B \propto 1/r$ and 
$\rho \propto 1/r^2$, one would then get acceleration until 
$\gamma_{wind} \approx \sigma_0$ and $\sigma (r) \rightarrow 1$. 

This expectation is not correct.  Acceleration continues until the wind {\em speed}
reaches the magnetosonic speed.  The acceleration is parallel to the poloidal field
and the poloidal velocity. The rate of change of the poloidal
speed is expressed by 
\begin{equation}
\rho \frac{d(\gamma v_p)}{dt} = \rho c^2 \beta_p \frac{\partial}{\partial r} 
 \frac{\beta_p}{\sqrt{1-\beta_p^2}} = 
      \rho c^2 \gamma^3 \beta_p \frac{\partial \beta_p}{\partial r}
 \sim \frac{\rho c^2 \gamma^3 \beta_p^2}{r} \nonumber
\end{equation}
- that is, the accelerating magnetic force must oversome the {\em longitudinal}
inertial mass/volume $\rho \gamma^3$ of the plasma, since the force is parallel
to the outflow velocity. Therefore, acceleration ceases once $\gamma$ rises
to $(B^2 /4\pi \rho c^2)^{1/3} = \sigma_0^{1/3}$, not to $\sigma_0$. This 
magnetosonic speed
based on the longitudinal mass  $v_{ms} = B/\sqrt{4\pi \rho \gamma^3 }$ has appeared
in all the formal discussions of relativistic wind outflow, always without
physical explanation.  Since this physics is generic to all outflow geometries, all
ideal MHD relativistic winds should have asymptotic 4-velocity limited to\footnote{In
magnetic geometries with field lines focused toward the magnetic axis, weak
logarithmic acceleration can occur beyond the radius where the velocity reaches
$v_{ms}$ (Begelman and Li 1994), but this effect of hoop stress does not have 
much practical significance.} $\gamma_\infty \sim \sigma_0^{1/3}$.

The location of the magnetosonic radius, located where $v(R_{ms}) = v_{ms}(R_{ms})$,
{\em is} sensitive to magnetic geometry. In the 1D flow models
(Michel 1969, Goldreich \& Julian 1970), $R_{ms} = \infty$ in cold flows.  
In multi-dimensional models, the inner regions of 
the flow are force free, with $\gamma \approx r/R_L$,
(Buckley 1977, Beskin et al. 1998, Contopoulos and Kazanas 2002, 
Arons, in preparation).  Such
``surf-riding'', with particles stuck to accelerating field lines, 
persists until $v = v_{ms}$, which, in monopole geometry, occurs at
$R_{ms} \approx \sigma_0^{1/3} R_L$; at larger
radii, the wind coasts, with\footnote{Contopoulos and Kazanas (2002) and 
Arons (2003a,b) suggested that acceleration in the
force free fields 
continues until equipartition, $\gamma \rightarrow \sigma_0$. This is wrong.}
 $\gamma \approx \sigma_0^{1/3}$. One can readily show that in magnetic
geometries with poloidal field lines flaring more rapidly than those of a monopole
even for a limited range of radii beyond the light cylinder that $R_{ms}$ moves
in to be close to the light cylinder (Arons, in preparation).

Thus ideal MHD wind theory leaves both high asymptotic $\gamma$ and low asymptotic
$\sigma$ quite unexplained, and will always be an inadequate theory to understand
the most elementary interpretation of the observations.

\section{Dissipation}

Arguments such as these suggest that dissipation of magnetic energy in the
asymptotic wind zone $R_L \ll r < R_{shock}$ must occur, in order to understand
the observed high $\gamma_1$ and low $\sigma_1$.  

The restriction of the observed low $\sigma$, high $\gamma$  flow to the rotational 
equator is an important clue. The equatorial zone is the likely location of a current
sheet separating opposite magnetic polarities in opposite hemispheres. Current sheets
are almost always sites of magnetic dissipation. In oblique rotators, this sheet
should be wrinkled, taking the form of a frozen in wave in a sector around the
rotational equator (Bogovalov 1999).
Possible dissipation mechanisms include some form of reconnection (Kirk and 
Skj{\ae}raasen 2003), dissipation of the wrinkles at the termination shock (Lyubarsky
2003), if they survive all the way to the shock, and instability of he 
current sheet with respect to the emission of strong electromagnetic waves
which propagate with respect to the wind (Arons 2003a, b), which rapidly
dissipate (Melatos 1998).  While all such mechanisms suggest substantial
photon emission, relativistic beaming means that unless the observer lies within 
the restricted rotational latitude range filled by the dissipative sheet
($|\lambda | < 10^\circ $ in the Crab), the photon signature of the dissipation would
be missed.  Finding a high voltage pulsar observed from close to the equatorial
plane would greatly improve our ability to model these phenomena.

\section{Acknowledgments}

I acknowledge assistance from NASA grants
NAG5-12031 and HST-AR-09548.01-A, the Miller Institute
for Basic Research in Science and the taxpayers of California.

\section{References}

Arons, J. 1998, Mem. Soc. Ast. It., 69, 989  \\
Arons, J. 2003a, Ap.J., 589, 871 \\
Arons, J. 2003b, Adv. in Space. Res., in press\\
Begelman, M.C., and Li, Z.-Y. 1994, Ap.J., 426, 269 \\
Beskin, V., Kuznetsova, I., and Rafikov, R. 1998, MNRAS, 299, 341 \\
Bogovalov, S. 1999, A\&A, 349, 1017  \\
Buckley, R. 1977, Nature, 266, 37 \\
Contopoulos, J., Kazanas, D. and Fendt, C. 1999, Ap.J., 511, 351 \\
Contopoulos, J., and Kazanas, D. 2002, Ap.J., 566, 336\\
Gallant, Y.A., van der Swaluw, E., Kirk, J.G., \& Achterberg, A. 2002, in \\
\hspace*{0.15in} `Neutron Stars and Supernova Remnants', P.O. Slane and \\
\hspace*{0.15in} B.M. Gaensler, eds. (San Francisco: ASP), 99 \\
Gaensler, B., Arons, J., Kaspi, V. et al. 2002, Ap.J., 569, 878 \\
Goldreich, P., \& Julian, W. 1970, Ap.J., 160, 971 \\
Hester, J.J., Mori, K., Burrows, D., et al. 2002, Ap.J., 577, L49 \\
Kennel, C.F., and Coroniti, F.C. 1984a, 283, 694; 1984b, 283, 710 \\
Kirk, J.G., and  Skj{\ae}raasen, O. 2003, Ap.J., 591, 366 \\
Lyubarsky, Y.E.  2003, MNRAS, 345, 153 \\
Melatos, A. 1998, Mem. Soc. Ast. It., 69, 1009 \\
Michel, F.C. 1969, Ap.J., 158, 727 \\
Michel, F.C. 1971, Comm. Ap., 3, 80 \\
Reynolds, S. 2003, in Proceedings of IAU Colloquium 192, \\
\hspace*{0.15in} 10 Years of SN1993J (San Francisco: ASP), in press (astro-ph/0308483)\\
Spitkovsky, A. \& Arons, J. 2003, Ap.J., in press

\end{document}